\documentclass{article}
\usepackage{spconf,amsmath,graphicx}

\usepackage{multirow}

\usepackage{xcolor}
\usepackage{soul}
\usepackage{adjustbox}

\title{Efficient dynamic filter for robust and low computational feature extraction}
\name{Donghyeon Kim$^1$, Jeong-gi Kwak$^1$, Hanseok Ko$^1$}
\address{
  $^1$Korea University, South Korea}
\copyrightnotice{978-1-6654-7189-3/22/\$31.00~\copyright2023 IEEE}
\begin{document}
%
\maketitle
\begin{abstract}
The unseen noise signal is difficult to anticipate, and various approaches have been developed to address this issue. In our earlier work, we proposed a lightweight dynamic filter by splitting the filter into kernel and spatial parts. This small footprint model showed robust results in an unseen noisy environment. However, a simple pooling process for dividing the feature would limit the performance. In this paper, we propose an efficient dynamic filter to enhance the performance of the existing dynamic filter. Instead of the simple feature mean, we separate the input features as non-overlapping chunks, and separable convolutions take place for each feature direction. We also propose a dynamic filter based attention pooling method. These methods are applied to the kernel part in our previous work, and experiments are carried out for keyword spotting and speaker verification. We confirm that our proposed method performs better in unseen environments than the recently developed models.
\end{abstract}
\section{Introduction}
{\let\thefootnote\relax\footnotetext{This work was supported by Korea Environment Industry \& Technology Institute(KEITI) through Exotic Invasive Species Management Program, funded by Korea Ministry of Environment(MOE) (2021002280004). Corresponding Author:Hanseok Ko.
}}

In audio-based deep learning applications, noise disturbance would occur performance degradation since the noise signal may shift the domain of target signals. In particular, unseen noise is difficult to anticipate and various methods have been developed to mitigate its effects. Through adversarial loss \cite{pascual2017segan,fu2019metricgan,kao2020orthogonal} and speech quality metric learning \cite{hu2007evaluation,fu2019learning,xu2021deep}, a generative model enhances the speech signal against noise. In addition, a data augmentation \cite{kim2021specmix} would help to enhance the speech signal and classification-driven approaches \cite{shon2019voiceid,kim2020dual} would mitigate noisy signals through an end-to-end fashion. Assuming that noise signals are cumbersome features for the classification task, a T-F mask is estimated from an input T-F feature and it is recursively multiplied with the input T-F feature to mitigate the noise effect. Although these methods perform reasonably well in various applications, their implementations are computationally expensive. In our previous work, we proposed a lightweight dynamic filter \cite{kim2021lightweight} as an alternative to robust and low resource keyword spotting. The filter was divided into two subtypes: Pixel Dynamic Filter (PDF) and Instance-Level Dynamic Filter (IDF). For PDF, a single channel convolution layer strides T-F features and the output of the convolution is considered pixel-level scalar weights. In IDF, weight vectors are extracted by using global averaging pooling and Fully Connected (FC) layers. These learning methods reduce model complexity and improve Keyword Spotting (KWS) performance in various noisy environments.

Although the feature averaging process would relieve the computational cost, a simple feature averaging (e.g. feature mean) might be difficult for capturing task salient features and it would degrade the performance of dynamic weights. It would be particularly weak in noisy environments that are unseen. To compat this issue, this paper propose Chunk Separated Convolution (CS-Conv) and Dynamic Attention Pooling (DAP) as effective dynamic filters that are robust against unseen signals. Based on the Dual-Path RNN framework \cite{9054266}, the T-F features are split into non-overlapped chunk features, and separable convolutions are performed on intra chunks and inter chunks, respectively. Additionally, our proposed DAP maps the output feature of CS-Conv to a temporal saliency-based embedding vector. A convolution layer in DAP produces vector weights that are used to compute attention scores depending on temporal frames. This process enables the dynamic filter to focus on temporally salient features to obtain the embedding vector.
The proposed method is applied to IDF of our previous work and experiments are carried out on Speaker Verification (SV) and KWS tasks. For KWS and SV evaluations, we use Speech command data \cite{warden2018speech} and voxceleb data \cite{nagrani2020voxceleb,chung2018voxceleb2}. We confirm that our proposed method enhances the performance of KWS and SV in unseen environments (unseen noise and unseen speakers) over our previous work with similar computation costs.
\section{Related Works}
\begin{figure}[t]
     \centering
     \includegraphics[scale=0.30]{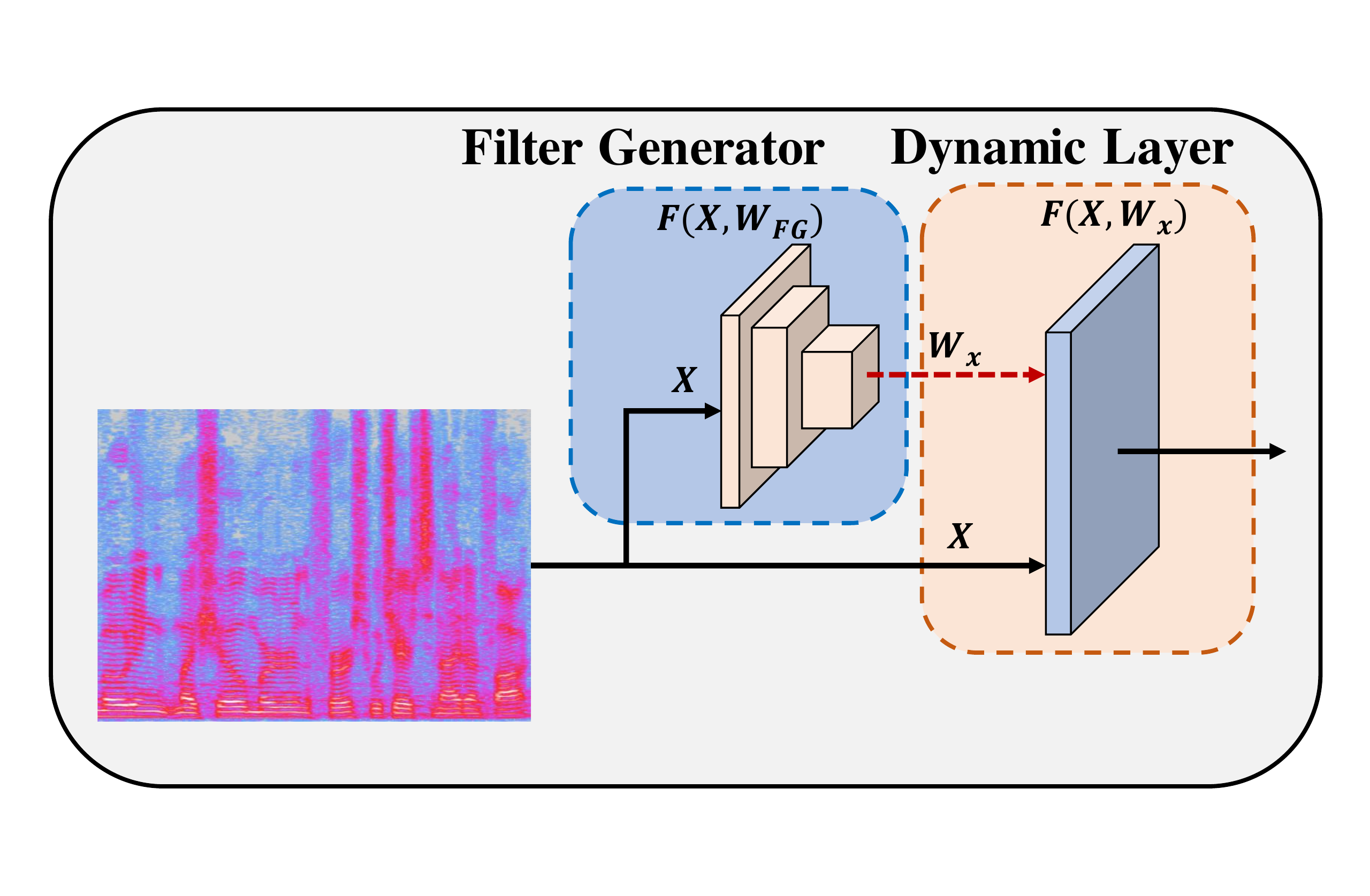}
     \caption{The pipeline of dynamic filter network:$F(X,W)$ denotes neural network implementation with an input feature ($X$) and weight vector ($W$). $W_{fg}$ denotes the weight of the filter generator model. $W_{x}$ denotes the output of the filter generator which is utilized as weights in the dynamic layer.}
    \label{fig1}
\end{figure}

Dynamic Filter Network \cite{jia2016dynamic} (DFN) is a filter adaptation method that dynamically modifies the filter values based on the input feature, and the process is depicted in Fig 1. Instead of optimizing the initialized weight vector, output features of the neural networks (filter generator) are used as weight vectors of the main task model (dynamic layer). In this learning process, weight vectors are produced by the manner of the filter generator and the weights are independently generated by each instance basis.
This property would enhance the flexibility, where the weights would not be frozen, and it has shown reasonably acceptable performance in various deep learning applications \cite{wu2019pay,kim2020dual,fujita2020attention}. However, the implementation of DFN requires a high computational cost (time and space complexity) when the input example has a lot of bases (time step or image patch) and the size of the input feature is usually large in modern deep learning applications. To address this issue, a small footprint dynamic model has been proposed. Instead of producing dynamic weights
for each patch unit, decoupled dynamic filter \cite{zhou2021decoupled} produces the weights as channel and spatial (kernel) units separately by splitting the filter generation network as two branches (channel and spatial parts). Based on the Global Average Pooling (GAP), the input features are divided into channel and spatial, and the dynamic weights are produced in two different models, respectively. The outputs of each model are then combined to perform the convolution.

\begin{figure}[t]
     \centering
     \includegraphics[scale=0.28]{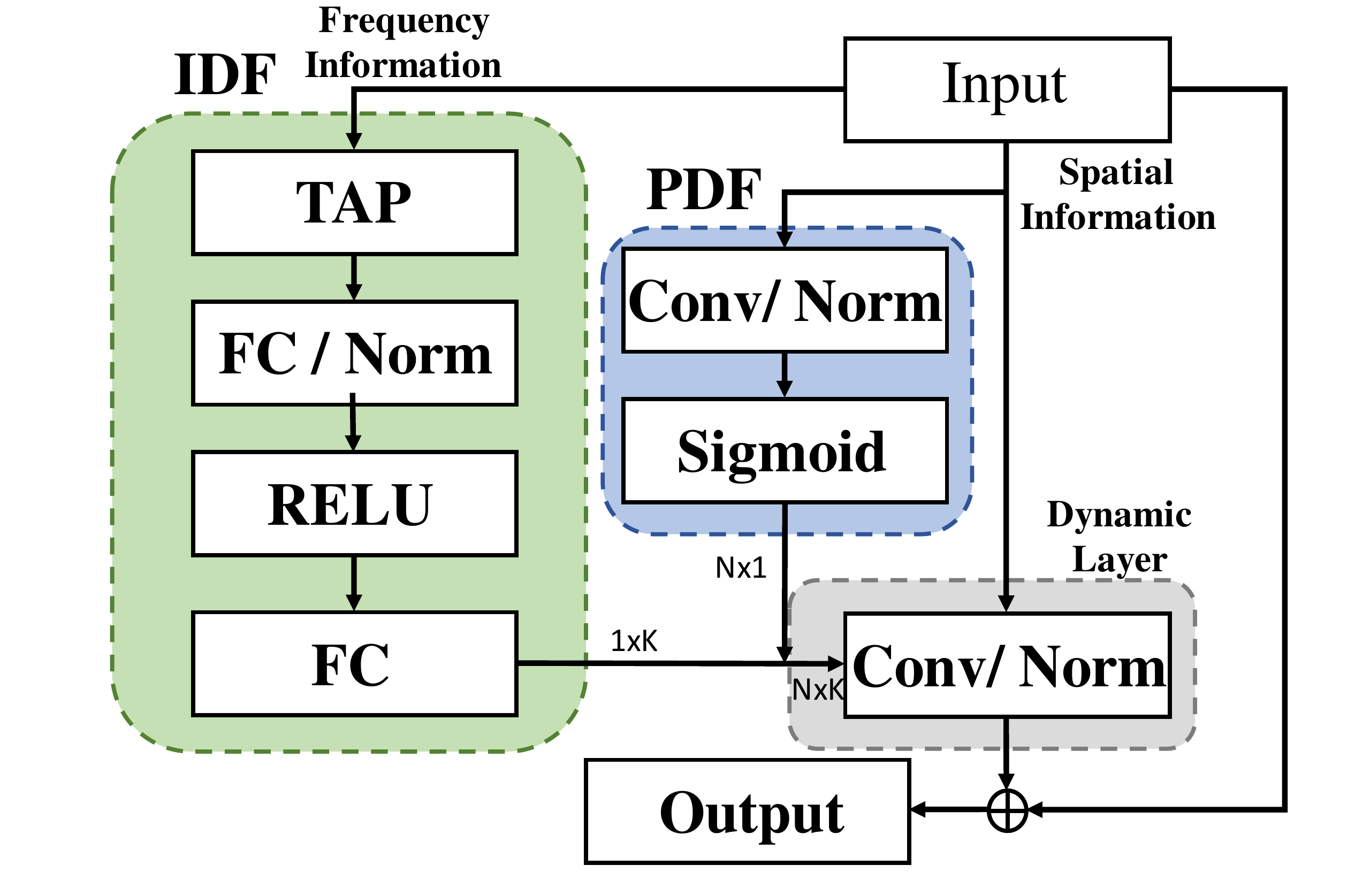}
     \caption{The pipeline of our previous work: PDF, IDF denote Pixel Dynamic Filter and Instance-level Dynamic Filter, respectively. $k$ and $N$ denote the size of kernel and pixel, respectively. The IDF produces $1 \times K$ weight and the PDF process produces $N \times 1$ weight. These weights are multiplied and the convolution is computed by using the generated weights}
    \label{fig4}
\end{figure}

In our earlier work, we proposed a front-end light dynamic filter for KWS, and the process is described in Fig 2. We divided the filter generator network into pixel and kernel weight models respectively. In the pixel model (PDF), a single layer of CNN and normalization processes produce pixel-level weights, which represent the scale of each patch. In the kernel weight model (IDF), temporal feature averaging and FC layers are employed to obtain the convolution kernel weights depending on each utterance example. These two weights are combined to perform single-channel CNN and it shows robust KWS results against unseen noise environments.
Although these learning frameworks reduce the computational cost of the DFN, their performance would be limited as feature averaging is employed to reduce the dimension of the feature. As a simple mean operation might generate biased weights, the performance becomes limited in a noisy environment. In this work, we aim to develop an efficient dynamic filter, where the filter generator's small footprint dynamic filter would work efficiently over a simple feature average.
\section{Efficient dynamic filter}
To mitigate the biased weight problem, we aim to develop an efficient feature mapping model which is fully trainable and low computation. To this end, we employ DP-RNN and attention mechanisms to develop the chunk separable convolution and dynamic attention pooling.
\subsection{Chunk Separable Convolution}
Our proposed CS-Conv is motivated by the Dual-Path Recurrent Neural Network (DP-RNN) \cite{9054266}. In DP-RNN, temporally overlapping T-F chunks are fed to two ways of RNNs (inter and intra chunk), and the output features are assembled by overlap adding. Instead of the RNN model, we utilized separable CNN by reducing the temporal size, and it is computationally more efficient than RNN. To this end, We divide T-F features into several chunks that do not overlap in time, and separable convolutions are applied to intra chunk and inter chunk, respectively. We change the format of the input T-F feature ($x \in R^{[F,T]}$, where $F$ and $T$ denote frequency and temporal dimension respectively) as non-overlapping chunks ($x_c \in R^{[F,C,T/C]}$, where $c$ denotes chunk size). Then, separable convolutions are conducted to the intra and the inter-chunk directions. In the intra chunk direction, $w_{itra}\in R^{[K,K,T/C]}$, where $k$ denotes kernel size, strides each chunk feature and the outputs are fed to the inter chunk convolution layer which has $w_{iter}\in R^{[K,K,C]}$ of CNN kernel. The CS-Conv procedure is as follows: 
\begin{equation}
   CSconv(x_c)= \sigma(conv(\sigma(conv(x_c,w_{itra})),w_{iter})),
  \label{eq1}
\end{equation}
where $\sigma (\cdot)$ and $conv (\cdot)$ denote the feature normalization and the operation of CNN, respectively. In each convolution layer, the kernels are put in stride to reduce the temporal dimension, and feature normalization is applied to the output of CNN. The intra and inter-chunk CNNs would stride various intervals of T-F patterns, and the output features might represent long-term and short-term traits with preserving the original frequency dimension.
\subsection{Dynamic Attention Pooling}

\begin{figure}[t]
     \centering
     \includegraphics[scale=0.28]{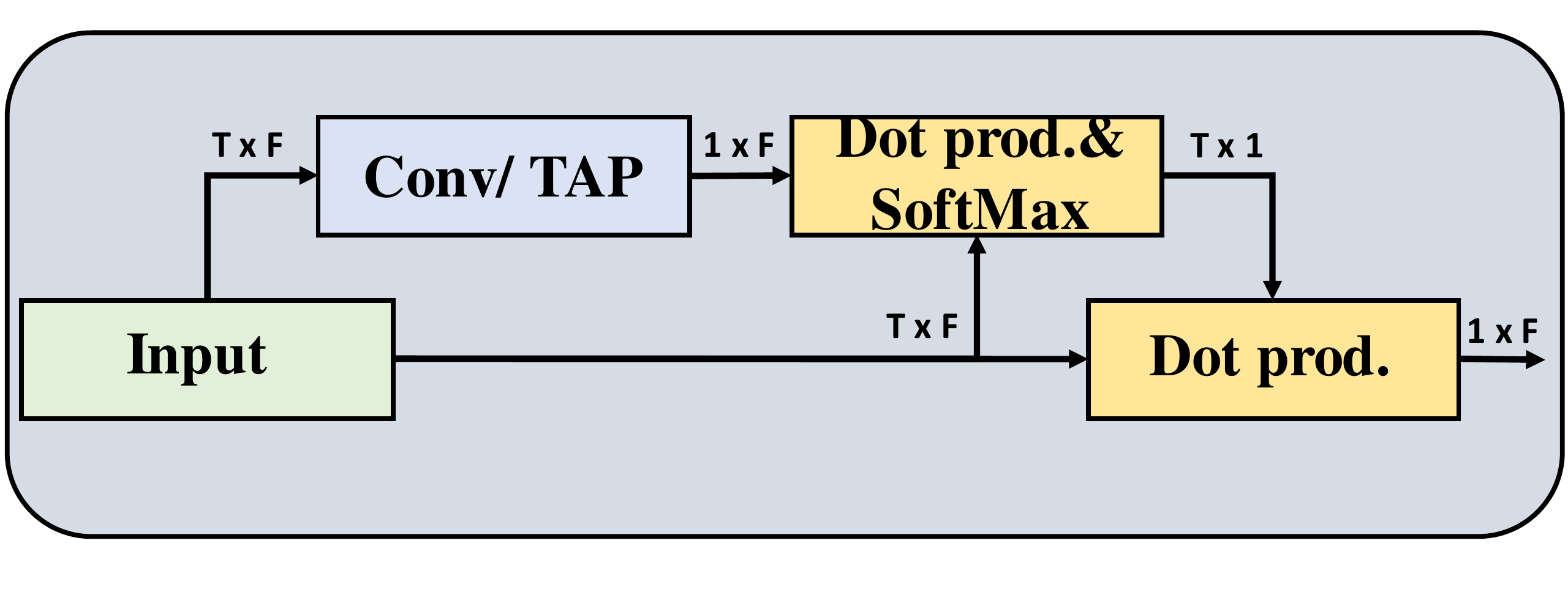}
     \caption{The pipeline of dynamic attention pooling: Dot prod. denotes the dot product computation.}
    \label{fig2}
\end{figure}

In the deep learning framework, GAP is generally applied to obtain a low-dimensional feature embedding. Based on this, mean-standard deviation pooling \cite{snyder2018x} and a Self Attention Pooling (SAP) \cite{safari2019self} were developed. In particular, SAP adjusts temporal saliency using the dot product between T-F features and learning variables. However, as SAP utilizes static weights, its performance might be limited. In our proposed DAP, the idea of DFN is employed to produce weights for SAP, and the process is described in Fig 3. The process of DAP is as follows: 
\begin{equation}
      w_d= TAP(conv(x,w)), 
  \label{eq2}
\end{equation}

\begin{equation}
      w^{Att}= \rho(x \times w_d), 
  \label{eq3}
\end{equation}

\begin{equation}
   DAP(x,w_d)= \sum_{i}^{T} w^{Att}_i  \times x_i, 
  \label{eq4}
\end{equation}
where $\rho(\cdot)$ denotes softmax normalization. A single layer of 1D convolution layer, which strides the temporal axis, is employed to produce a temporally low-resolution feature, and Temporal Average Pooling (TAP) is conducted to obtain dynamic weights ($w_d \in R^{[F,1]}$). Then, the weights and the input features are multiplied to compute the temporal weights, and the softmax normalization across the temporal step is performed. Finally, the attention scores ($w^{Att} \in R^{[1,T]}$) are multiplied with the input features to obtain low-dimensional feature embedding.
\subsection{Network Architecture}
\begin{figure}[t]
     \centering
     \includegraphics[scale=0.28]{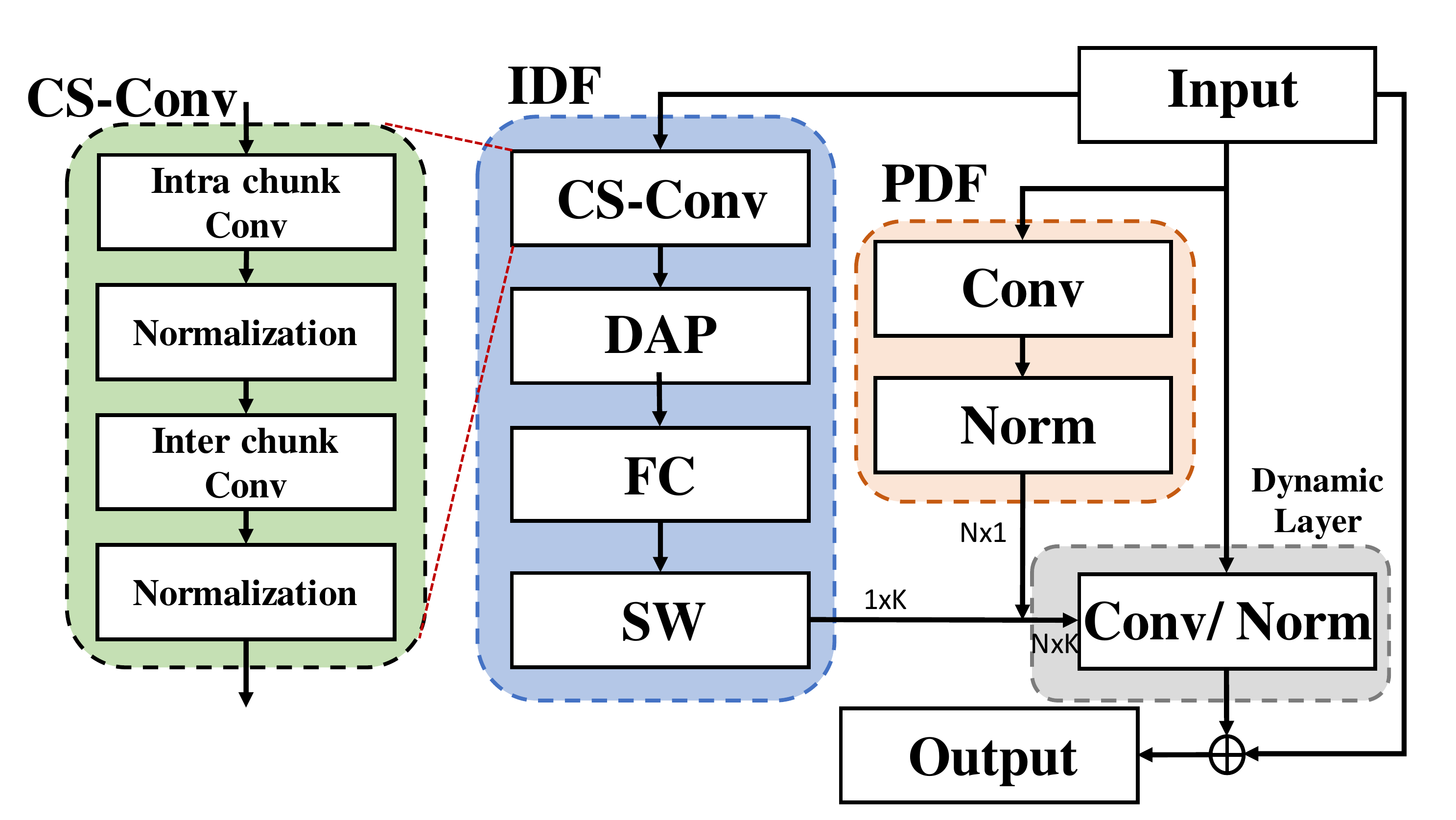}
     \caption{The pipeline of our proposed method: PDF, IDF, and CS-Conv denote Pixel Dynamic Filter, Instance-level Dynamic Filter, and Chunk Separable Convolution, respectively. $k$ and $N$ denote the size of kernel and pixel, respectively. The IDF produces $1 \times K$ weight and the PDF process produces $N \times 1$ weight. These weights are multiplied and the convolution is computed by using the generated weights.}
    \label{fig3}
\end{figure}

As shown in Fig 4, we modify the IDF part from our previous work in Fig 2. Instead of simple pooling and FC layers, the CS-Conv block is utilized to compute long-term and short-term features, and then DAP is applied to map low-dimensional embedding. Then, an FC layer and Swish activation \cite{ramachandran2017searching} are used to generate a weight vector. Our dynamic filter and main task model are both optimized by classification loss (cross-entropy loss). For feature normalization, we use the Instance Normalization (IN) \cite{ulyanov2016instance} framework, where the mean-standard deviation normalization and learnable affine transform are applied to the frequency axis. With CS-Conv, we apply a feature normalization to each chunk individually. In the other normalization layers, we divide the input features into two temporal chunks, and apply feature normalization to each chunk.
\subsection{Learning details}
In CS-Conv, we use $2 \times 2$ of CNN kernels with a dilation size of 2 and stride size of (1,2) in both convolution layers. In DAP, $1 \times 25$ CNN kernel with stride size of (1,10) is utilized to generate the dynamic weights, which is a 40-dimensional vector. The output of DAP is fed to $40 \times 9$ of FC layer to produce the IDF weighs. In PDF, $3 \times 3$ kernel with a dilation size of 2 is used to produce the PDF weights. 
\section{Keyword Spotting}
\subsection{Experimental setup}
{\bf Dateset.} We used the speech command datasets v1 and v2 \cite{warden2018speech} for evaluating KWS performance and followed the same data protocol in the DB guideline. We utilized 10 keywords with two extra classes (unknown or silent) for model training and injected background noise with random time-shifting. For evaluating robustness against noise, we utilized the DCASE \cite{mesaros2019acoustic}, Urbansound8K \cite{salamon2014dataset} and WHAM \cite{Wichern2019WHAM} datasets. We randomly selected an audio sample from the noise data and mixed it with the test data with 5 SNRs [20dB, 15dB, 10dB, 5dB, and 0dB].\\
{\bf Implementation detail.} 40 coefficients of Mel Frequency Cepstral are constructed with 30ms of windows, 10ms overlap, and 64 Mel-filters. This process gives a (40,98) size of audio features. In the training process, we used a batch size of 100, 30k iterations, and an ADAM optimizer with a 0.001 initial learning rate. Every 10K iteration, the learning rate is decreased by 0.1 rates. In the PDF and the dynamic convolution process, we used $3\times3$ CNN kernel ($k=9$) dilated by 2 with stride=1. In the IDF, $2\times2$ kernel dilated by 2 with stride of [1,2] is utilized to each CSconv and FC layer follows $40 \times k$ dimension. We do not use Spec-Augmentation \cite{park2019specaugment} for fair comparison. 
\subsection{Baselines}
Four different baseline architectures are used for comparisons. We applied our method to the front-end of the TENet12 architecture \cite{li2020small} and compared the KWS result with the following baselines.\\
{\bf TCNet.} TCNet \cite{choi2019temporal} (or TC-Resnet) utilizes temporal convolution for low computational cost model. TCNet8 contains 3 convolution blocks and 1 FC layer. Each convolution block has two layers of temporal convolution with a skip-connection. Similarly, TCNet14 contains 6 convolution blocks and 1 FC layer. \\
{\bf TENet.} TENet \cite{li2020small} utilizes depth-separable temporal convolution. A convolution block contains two 1D convolution layers and a single separable convolution layer. TENet6 and TENet12 has 6 and 12 convolution blocks respectively. TENet has 32 output channels for each convolution block, and TENet-n has 16 output channels for each convolution block.\\
{\bf MHA-RNN.} MHA-RNN\cite{rybakov2020streaming} utilizes CRNN and self-attention models to compute the keyword embedding vector. Two layers of FC produce KWS probabilities, and spec augmentation \cite{park2019specaugment} is used to train the model.\\ 
{\bf Neural Architecture Search.} Neural Architecture Search (NAS) is a network designing method that minimizes search costs (Flops, memory, accuracy, etc.). We compared our method with several NAS methods including Differentiable Architecture Search (DARTS). Please see details of the model in \cite{mo2020neural,zhang2021autokws}.\\
{\bf Lightweight convolution.} Lightweight Convolution \cite{wu2019pay} (LConv.) block is designed to perform low computation separable convolution through weight sharing and weight normalization. It contains two layers of linear, a Gated Linear Unit (GLU), and the separable convolution. We use the single LConv block in the front end of the TENet12 model. In the first linear layer, the frequency dimension of the T-F features is increased by 80, and the GLU activation is applied. Then, the separable convolution with $H=10$ and the other linear layer is performed by preserving the feature dimension. Additionally, instead of employing static weights in the separable convolution, we utilized a single linear layer to produce the weight of the convolution (Dyconv).\\
{\bf LDy-TENet.} We previously developed LDy-TENet \cite{kim2021lightweight} which applied a lightweight dynamic filter to the front-end of the TENET models. In the PDF process, $3\times3$ kernel with a dilation size of 2 is used. In IDF process $40 \times 40$ and $40 \times k$ dimension of weights are used.
 \subsection{Results and discussion}
\begin{table}[t]
\scriptsize
  \centering
  \label{tab2}
  \begin{tabular}{|c|c|cc|cc|}
\hline
    \multicolumn{1}{|c|}{\multirow{2}{*}{\textbf{Model}}}&\multicolumn{1}{c}{\multirow{2}{*}{\textbf{(Par.,Flops.)}}}&\multicolumn{2}{|c|}{V1}&\multicolumn{2}{c|}{V2}\\\cline{3-6}
    \multicolumn{1}{|c|}{}&\multicolumn{1}{c|}{}&\textbf{Acc}& \textbf{Best}& \textbf{Acc}& \textbf{Best}\\
\hline
    TCNet8\cite{choi2019temporal}& (145K,4.40M)& - &96.2&- &- \\
    TCNet14\cite{choi2019temporal}& (305K,8.26M)& - &96.6&96.53&96.8 \\
\hline
    TENet6-n\cite{li2020small}& (17K,1.26M)& - &96.0& 96.30&96.5 \\
    TENet12-n\cite{li2020small}& (31K,1.97M)& - &96.3& 96.55&96.9 \\
    TENet6\cite{li2020small}& (54K,3.95M)& - &96.4& 96.83&97.0 \\
    TENet12\cite{li2020small}& (100K,6.42M)& - &96.6 &97.10&97.3\\
\hline
    MHA-RNN$^{\dagger}$\cite{rybakov2020streaming}& (743K,87.2M)& - &97.2& -&98.0 \\
\hline
    NAS2\cite{mo2020neural}& (886K,-)& - &97.2& -& -\\
    Random\cite{zhang2021autokws}& (196K,8.8M)& 96.58 &96.8& -&- \\
    DARTS\cite{zhang2021autokws}& (93K,4.9M)& 96.63 &96.9& 96.92&97.1 \\
    F-DARTS\cite{zhang2021autokws}& (188K,10.6M)& 96.70 &96.9&97.11& 97.4\\
    N-DARTS\cite{zhang2021autokws}& (109K,6.3M)& 96.79 &97.2& 97.18&97.4 \\
\hline
    LightConv\cite{wu2019pay}&(105K,7.40M)&96.88&97.0& 97.24&97.3\\
    DyConv\cite{wu2019pay}&(107K,7.69M)&96.89&97.1& 96.26&97.4\\
\hline
    LDy-TENet6-n\cite{kim2021lightweight}& (19K,1.48M)& 96.48&97.0 & 96.10&96.8\\
    LDy-TENet12-n\cite{kim2021lightweight}& (33K,2.19M)& 96.69&96.9& 96.80&97.2 \\
    LDy-TENet6\cite{kim2021lightweight}& (56K,4.17M)& 96.77&96.9 & 97.26&97.4\\
    LDy-TENet12\cite{kim2021lightweight}& (102K,6.64M)& 96.95 &97.1 & 97.35&97.6\\
\hline
    EDy-TENet12 (Ours)& (102K,6.68M)& \textbf{97.07} &\textbf{97.4}& \textbf{97.42}&\textbf{97.8} \\
\hline
  \end{tabular}
    \caption{Comparison with lightweight models on Speech Command v1 and v2: Notation of $\dagger$ denotes the application of Spec-Augmentation \cite{park2019specaugment}. For an accurate experiment, 8 times averaging accuracy and the best performance are presented.}
\end{table}

\begin{table}[t]
  \label{tab3}
  \centering
  \begin{tabular}{|c|ccc|}
\hline
    \textbf{Acc}& \textbf{GAP}& \textbf{SAP}&\textbf{DAP}\\
\hline
    \textbf{V1}& 97.02&97.04&\textbf{97.07}\\
    \textbf{V2}& 97.22&97.35&\textbf{97.42}\\
\hline
  \end{tabular}
  \caption{Comparison of feature pooling methods depending on the Speech Commands datasets v1 and v2.}
\end{table}

\begin{table*}[t]
  \label{tab4}
  \centering
  \normalsize
\begin{tabular}{|c|c|cccccccccccc|}
\hline
\multicolumn{1}{|c|}{\multirow{3}{*}{\textbf{Noise}}} & \multicolumn{1}{|c|}{\multirow{3}{*}{\begin{tabular}[c]{@{}c@{}}SNR\\ (dB)\end{tabular}}} & \multicolumn{12}{c|}{\textbf{Model}}\\ \cline{3-14}
\multicolumn{1}{|c|}{}&\multicolumn{1}{c|}{}&\multicolumn{2}{c}{\textbf{Ours}}&\multicolumn{2}{c}{\textbf{LDy}}&\multicolumn{2}{c}{\textbf{Dconv}}&\multicolumn{2}{c}{\textbf{Lconv}}& \multicolumn{2}{c}{\textbf{TENet}}& \multicolumn{2}{c|}{\textbf{TCNet}}\\\cline{3-14}
\multicolumn{1}{|c|}{}&\multicolumn{1}{c|}{}&\textbf{v1}&\textbf{v2}&\textbf{v1}&\textbf{v2}&\textbf{v1}& \textbf{v2}&\textbf{v1}&\textbf{v2}&\textbf{v1}&\textbf{v2}&\textbf{v1}&\textbf{v2}\\
\hline
\multirow{5}{*}{{\begin{tabular}[c]{@{}c@{}}\textbf{DCASE}\\\cite{mesaros2019acoustic}\end{tabular}}}                                                                                 &\textbf{20} &\textbf{96.72}& \textbf{97.04}& 96.58& 96.95&96.55& 96.74&96.62& 96.81&96.28& 96.59&95.85&96.09\\
						         &\textbf{15} &\textbf{96.50}& \textbf{96.45}&96.44& 96.31&96.35& 96.08&96.37& 96.08&96.23& 96.01&95.59&95.39\\
					            &\textbf{10} & \textbf{95.32}& \textbf{95.64}&95.10& 95.56&95.01& 95.09&95.05& 95.17&94.94& 95.15&94.25&94.33\\
                                &\textbf{5} & \textbf{93.54}&\textbf{93.42} &93.00& 93.35&92.27& 92.42&92.80& 92.80&92.87& 92.53&91.79&91.74\\
					            &\textbf{0} & \textbf{89.36}& \textbf{88.73}&88.41& 88.46&87.51& 86.64&87.67& 87.01&88.14& 86.65&86.01&85.96\\\hline
\multirow{5}{*}{\begin{tabular}[c]{@{}c@{}}\textbf{Urban}\\\cite{salamon2014dataset}\end{tabular}}                                                                                    &\textbf{20} & \textbf{96.08}& \textbf{96.35}&95.94& 96.33&95.81& 95.81&95.84& 96.06&95.72& 95.92&95.16&95.08\\
						         &\textbf{15} & \textbf{95.20}& \textbf{94.95}&95.09& 94.91&94.69& 94.69&94.49& 94.45&94.69& 94.31&93.61&93.53\\
					            &\textbf{10} & \textbf{93.10}& \textbf{93.04}&92.82& 92.79&92.03& 92.03&92.25& 91.97&92.17&92.07 &90.77&90.81\\
                            &\textbf{5} & \textbf{89.63}& 87.51&89.19& \textbf{87.56}&87.80& 87.08&87.60& 85.87&87.97&86.18 &86.38&84.53\\
					            &\textbf{0} & \textbf{80.50}& \textbf{79.63}&78.55& 79.38&76.20& 75.86&76.26& 76.15&77.54& 76.44&74.20&74.34\\\hline
\multirow{5}{*}{\begin{tabular}[c]{@{}c@{}}\textbf{WHAM}\\\cite{Wichern2019WHAM}\end{tabular}}                                                                                        &\textbf{20} & \textbf{96.17}& \textbf{96.30}&96.09& 96.27&95.96& 96.08&96.02& 96.13&95.75& 95.99&95.43&95.35\\
						         &\textbf{15} & \textbf{95.50}& \textbf{95.40}&95.47& \textbf{95.40}&96.12& 94.61&95.23& 94.84&95.12& 94.74&94.44&93.45\\
					            &\textbf{10} & \textbf{93.06}& 93.39&92.91& \textbf{93.44}&92,39& 92.25&92.69& 92.71&92.67& 92.53&91.46&91.47\\
                                &\textbf{5} & \textbf{89.10}& \textbf{88.10}&88.19& 87.95&87.36& 86.26&87.83& 86.60&87.88& 86.37&85.72&85.42\\
					            &\textbf{0} & \textbf{77.79}& \textbf{76.97}&75.92& 76.37&74.52& 73.80&75.18& 74.30&75.19& 74.05&73.15&72.90\\\hline
\multicolumn{2}{|c|}{\textbf{Total-AVG.}} & \textbf{91.87}& \textbf{91.58}&91.31& 91.40&90.64& 90.25&90.79& 90.45&90.87& 90.37&89.59&89.36\\
\hline
\end{tabular}
  \caption{Comparison with Unseen noise environment on the Speech Commands datasets v1 and v2: experiments are performed on EDy-TENet12 LDy-TENet12, TENet12, Lconv, Dconv, and TCNet14 model. SNR denotes signal-to-noise ratio.}
\end{table*}




Tables 1, 2, and 3 summarize the KWS results on the Speech Command datasets, v1 and v2. The proposed method (EDy-TENet12) is applied to the IDF part of \cite{kim2021lightweight}. For a more thorough model evaluation, we repeated the experiment eight times.\\ 
{\bf Small footprint KWS.} Table 1 compares small footprint KWS performances with baseline methods. "Par." and "Flops." denote the number of training parameters and the computational cost, respectively. The results indicate that our proposed model improves the performance of the two datasets over the LDy model. While the number of parameters is almost the same, 0.04M of Flops increased. Particularly our method and the MHA-RNN method show similar results. Given that the computational cost of the MHA-RNN is much higher than our method, the performance improvement, which seems to be barely significant, would be considered a meaningful result. In Table 2, we compare our method with other feature pooling methods (GAP and SAP) to verify its effectiveness. The GAP, SAP, and DAP are compared based on EDy-TENet12. The SAP computes the temporal scores based on a 40-dimensional static weight vector. We confirm that the DAP shows a small improvement over the SAP that utilizes static weights.\\ 
{\bf Unseen noisy environment.} Table 3 summarizes the KWS results on the unseen noisy environments. These results indicate that our proposed model is more robust than the LDy and Dconv models that use a dynamic filter. Compared with LDy-TENet, our method takes 0.56\% (v1) and 0.18\% (v2) of improvement in performance on average. Even though performance improvement is not significant in the v2 dataset, v1 shows reasonably acceptable performance over the LDy-TENet12 model in terms of computational cost. Particularly 2\% performance improvement is attained in the v1 dataset of the Urban 0dB condition.\\
In summary, our method requires 1.5k parameters with 257k Flops to implement, whereas our previous work used 2k parameters with 220k Flops to implement. Given the computational cost, we can conclude that our method efficiently enhances the performance of KWS over our previous work and the other dynamic filter approach with limited resources.
\section{Speaker Verification}
\subsection{Experimental setup}
{\bf Dateset.} The speaker model is trained on the VoxCeleb2 dataset, and it is evaluated by the VoxCeleb1 data. We use 2 seconds of speech segments which are randomly chosen from each utterance, and 40 dimensions of the log-mel spectrogram are extracted by a window of 25ms and step of 10ms. We perform mean and variance normalization by instance normalization fashion. Also, we do not apply data augmentation and Voice Activation Detection (VAD) during the model training and evaluation.\\
{\bf Computational setup.} Based on the TENet12 and ECAPA-TDNN \cite{desplanques2020ecapa}, we compare our the performance of front-end models. We modify the final convolution block of the TENet12 model as 128 channels, and the vector format of the features is extracted by the TAP process. Then, we use two FC layers to match the output size to the number of speaker classes. A weights vector of $128 \times 512$ is used in the first FC layer, and a weights vector of $512 \times 5994$ is used to extract the probabilities for speaker identification. We utilize the output of the first FC layer as speaker embedding for each utterance. In the ECAPA-TDNN, we use 512 as the channel size of convolution layers and 256 dimensional speaker embedding is used to evaluate the performance.\\
{\bf Implementation details.} In the TENet training process, we use a batch size of 400, 500 epochs, and an ADAM optimizer with a 0.001 initial learning rate. In the ECAPA-TDNN training, we utilize a batch size of 50, 100 epochs. In each epoch, 500 utterances of samples are randomly selected from each speaker, and the learning rate is decreased by 0.999. The training takes approximately four days using the PyTorch package with RTX-3090 GPU. For the speaker model training, we use Angular Prototypical loss \cite{chung2020defence} with an utterance of each mini-batch as 2. We select ten samples of 2-second audio segments per test utterance with fixed intervals and compute the cosine-based similarity between all combinations (100 pairs) to evaluate the trained network. The averaging similarity of all possible pairs is used as the similarity score. We utilize Equal Error Rate (EER) and Minimal Detection Cost Function (Min. DCF) as the evaluation metrics.\\
{\bf Evaluation protocol.} In this work, two different evaluation protocols (Voxceleb1-test and Voxceleb1-H) are used. Voxceleb1-test consists of 40 speaker identities which are equal to the original test set in Voxceleb1. Voxceleb1-H contains 1251 speaker identities, and the sample pairs have the same nationality and gender. It is noted that the development set of VoxCeleb2 does not contain the same speaker identities as VoxCeleb1. This means that the evaluation is carried out in an open environment (unseen speaker).
\subsection{Result discussion}
Table 4 shows the SV results depending on the two different evaluation
protocols (Voxceleb1-test and Voxceleb1-H). We confirm that the dynamic filter in the front-end enhances the SV performance. Particularly our proposed model takes 0.53\% and 0.31\% of EER improvements in the Voxceleb1-H test set. As the weights of CNN are generated by the filter generator, it shows robust performance over unseen speakers. Especially our proposed method efficiently extracts feature, and it would enhance the performance of the dynamic filter better than our previous work.
\begin{table}[t]
  \label{tab1}
  \centering
    \normalsize
  \begin{tabular}{|c|c|cccc|}
\hline
    \multicolumn{1}{|c|}{\multirow{2}{*}{\textbf{SV}}}&\multicolumn{1}{|c|}{\multirow{2}{*}{\textbf{Front}}}&\multicolumn{2}{c}{Vox-test}&\multicolumn{2}{c|}{Vox-H}\\\cline{3-6}
\multicolumn{1}{|c|}{}&\multicolumn{1}{|c|}{}&\textbf{EER}& \textbf{M. DCF}&\textbf{EER}&\textbf{M. DCF}\\
\hline
\multirow{3}{*}{\begin{tabular}[c]{@{}c@{}}\textbf{TENet}\\\cite{li2020small}\end{tabular}}&
    \textbf{EDy}& \textbf{4.79}&\textbf{0.36}&\textbf{9.77}&\textbf{0.59}\\
    &\textbf{LDy}& 4.97&0.38&10.30&0.62\\
    &\textbf{Base}& 5.30&0.38&10.43&0.63\\
\hline

\multirow{3}{*}{\begin{tabular}[c]{@{}c@{}}\textbf{ECAPA.}\\\cite{desplanques2020ecapa}\end{tabular}}&
    \textbf{EDy}& \textbf{2.90}&\textbf{0.24}&\textbf{6.16}&\textbf{0.41}\\
    &\textbf{LDy}& 2.99&0.24&6.47&0.42\\
    &\textbf{Base}& 3.06&0.25&6.71&0.44\\
\hline

  \end{tabular}
  \caption{Comparison of the speaker verification results.}
\end{table}
\section{Conclusion}
In this paper, we proposed an efficient dynamic filter to extract robust and low computational features. In our earlier work, we showed that a dynamic neural architecture in the front-end of the model would enhance the robustness against unseen noise. However, as the simple feature means was applied to reduce the computational cost, the performance of the dynamic filter became limited, and it would particularly reduce performance in an unseen environment. To combat this limitation, we staged the separable convolutions using temporal chunks (intra and inter-chunks) to extract local and global features. Then, a dynamic filter-based feature pooling was employed to extract the low dimensional feature vector. This approach was applied to the Instance-level Dynamic Filter in our previous work, and performed relevant experiments on the keyword spotting and speaker verification tasks. From the experimental results, we verified that the proposed method enhances robustness against unseen environments (noise and speaker) compared to the recently developed methods with low computational cost.
\bibliographystyle{IEEEbib}
\bibliography{strings,refs}

\begin{thebibliography}{10}

\bibitem{pascual2017segan}
Santiago Pascual, Antonio Bonafonte, and Joan Serra,
\newblock ``Segan: Speech enhancement generative adversarial network,''
\newblock {\em arXiv preprint arXiv:1703.09452}, 2017.

\bibitem{fu2019metricgan}
Szu-Wei Fu, Chien-Feng Liao, Yu~Tsao, and Shou-De Lin,
\newblock ``Metricgan: Generative adversarial networks based black-box metric
  scores optimization for speech enhancement,''
\newblock in {\em International Conference on Machine Learning}. PMLR, 2019,
  pp. 2031--2041.

\bibitem{kao2020orthogonal}
Chao-Yuan Kao, Sangwook Park, Alzahra Badi, David~K Han, and Hanseok Ko,
\newblock ``Orthogonal gradient penalty for fast training of wasserstein gan
  based multi-task autoencoder toward robust speech recognition,''
\newblock {\em IEICE TRANSACTIONS on Information and Systems}, vol. 103, no. 5,
  pp. 1195--1198, 2020.

\bibitem{hu2007evaluation}
Yi~Hu and Philipos~C Loizou,
\newblock ``Evaluation of objective quality measures for speech enhancement,''
\newblock {\em IEEE Transactions on audio, speech, and language processing},
  vol. 16, no. 1, pp. 229--238, 2007.

\bibitem{fu2019learning}
Szu-Wei Fu, Chien-Feng Liao, and Yu~Tsao,
\newblock ``Learning with learned loss function: Speech enhancement with
  quality-net to improve perceptual evaluation of speech quality,''
\newblock {\em IEEE Signal Processing Letters}, vol. 27, pp. 26--30, 2019.

\bibitem{xu2021deep}
Ziyi Xu, Maximilian Strake, and Tim Fingscheidt,
\newblock ``Deep noise suppression with non-intrusive pesqnet supervision
  enabling the use of real training data,''
\newblock {\em arXiv preprint arXiv:2103.17088}, 2021.

\bibitem{kim2021specmix}
Gwantae Kim, David~K Han, and Hanseok Ko,
\newblock ``Specmix: A mixed sample data augmentation method for training
  withtime-frequency domain features,''
\newblock {\em arXiv preprint arXiv:2108.03020}, 2021.

\bibitem{shon2019voiceid}
Suwon Shon, Hao Tang, and James Glass,
\newblock ``Voiceid loss: Speech enhancement for speaker verification,''
\newblock {\em Proc. Interspeech}, pp. 2888--2892, 2019.

\bibitem{kim2020dual}
Donghyeon Kim, Jaihyun Park, David~K Han, and Hanseok Ko,
\newblock ``Dual stage learning based dynamic time-frequency mask generation
  for audio event classification,''
\newblock {\em Proc. Interspeech 2020}, pp. 836--840, 2020.

\bibitem{kim2021lightweight}
Donghyeon Kim, Kyungdeuk Ko, Jeong-gi Kwak, David~K Han, and Hanseok Ko,
\newblock ``Lightweight dynamic filter for keyword spotting,''
\newblock {\em arXiv preprint arXiv:2109.11165}, 2021.

\bibitem{9054266}
Yi~Luo, Zhuo Chen, and Takuya Yoshioka,
\newblock ``Dual-path rnn: Efficient long sequence modeling for time-domain
  single-channel speech separation,''
\newblock in {\em ICASSP 2020 - 2020 IEEE International Conference on
  Acoustics, Speech and Signal Processing (ICASSP)}, 2020, pp. 46--50.

\bibitem{warden2018speech}
Pete Warden,
\newblock ``Speech commands: A dataset for limited-vocabulary speech
  recognition,''
\newblock {\em arXiv preprint arXiv:1804.03209}, 2018.

\bibitem{nagrani2020voxceleb}
Arsha Nagrani, Joon~Son Chung, Weidi Xie, and Andrew Zisserman,
\newblock ``Voxceleb: Large-scale speaker verification in the wild,''
\newblock {\em Computer Speech \& Language}, vol. 60, pp. 101027, 2020.

\bibitem{chung2018voxceleb2}
Joon~Son Chung, Arsha Nagrani, and Andrew Zisserman,
\newblock ``Voxceleb2: Deep speaker recognition,''
\newblock {\em arXiv preprint arXiv:1806.05622}, 2018.

\bibitem{jia2016dynamic}
Xu~Jia, Bert De~Brabandere, Tinne Tuytelaars, and Luc~V Gool,
\newblock ``Dynamic filter networks,''
\newblock in {\em Advances in neural information processing systems}, 2016, pp.
  667--675.

\bibitem{wu2019pay}
Felix Wu, Angela Fan, Alexei Baevski, Yann~N Dauphin, and Michael Auli,
\newblock ``Pay less attention with lightweight and dynamic convolutions,''
\newblock {\em arXiv preprint arXiv:1901.10430}, 2019.

\bibitem{fujita2020attention}
Yuya Fujita, Aswin~Shanmugam Subramanian, Motoi Omachi, and Shinji Watanabe,
\newblock ``Attention-based asr with lightweight and dynamic convolutions,''
\newblock in {\em ICASSP 2020-2020 IEEE International Conference on Acoustics,
  Speech and Signal Processing (ICASSP)}. IEEE, 2020, pp. 7034--7038.

\bibitem{zhou2021decoupled}
Jingkai Zhou, Varun Jampani, Zhixiong Pi, Qiong Liu, and Ming-Hsuan Yang,
\newblock ``Decoupled dynamic filter networks,''
\newblock in {\em Proceedings of the IEEE/CVF Conference on Computer Vision and
  Pattern Recognition}, 2021, pp. 6647--6656.

\bibitem{snyder2018x}
David Snyder, Daniel Garcia-Romero, Gregory Sell, Daniel Povey, and Sanjeev
  Khudanpur,
\newblock ``X-vectors: Robust dnn embeddings for speaker recognition,''
\newblock in {\em 2018 IEEE International Conference on Acoustics, Speech and
  Signal Processing (ICASSP)}. IEEE, 2018, pp. 5329--5333.

\bibitem{safari2019self}
Pooyan Safari and Javier Hernando,
\newblock ``Self multi-head attention for speaker recognition,''
\newblock {\em arXiv preprint arXiv:1906.09890}, 2019.

\bibitem{ramachandran2017searching}
Prajit Ramachandran, Barret Zoph, and Quoc~V Le,
\newblock ``Searching for activation functions,''
\newblock {\em arXiv preprint arXiv:1710.05941}, 2017.

\bibitem{ulyanov2016instance}
Dmitry Ulyanov, Andrea Vedaldi, and Victor Lempitsky,
\newblock ``Instance normalization: The missing ingredient for fast
  stylization,''
\newblock {\em arXiv preprint arXiv:1607.08022}, 2016.

\bibitem{mesaros2019acoustic}
Annamaria Mesaros, Toni Heittola, and Tuomas Virtanen,
\newblock ``Acoustic scene classification in dcase 2019 challenge: closed and
  open set classification and data mismatch setups,''
\newblock in {\em Workshop on Detection and Classification of Acoustic Scenes
  and Events}, 2019.

\bibitem{salamon2014dataset}
Justin Salamon, Christopher Jacoby, and Juan~Pablo Bello,
\newblock ``A dataset and taxonomy for urban sound research,''
\newblock in {\em Proceedings of the 22nd ACM international conference on
  Multimedia}, 2014, pp. 1041--1044.

\bibitem{Wichern2019WHAM}
Gordon Wichern, Joe Antognini, Michael Flynn, Licheng~Richard Zhu, Emmett
  McQuinn, Dwight Crow, Ethan Manilow, and Jonathan Le~Roux,
\newblock ``Wham!: Extending speech separation to noisy environments,''
\newblock in {\em Proc. Interspeech}, Sept. 2019.

\bibitem{park2019specaugment}
Daniel~S Park, William Chan, Yu~Zhang, Chung-Cheng Chiu, Barret Zoph, Ekin~D
  Cubuk, and Quoc~V Le,
\newblock ``Specaugment: A simple data augmentation method for automatic speech
  recognition,''
\newblock {\em arXiv preprint arXiv:1904.08779}, 2019.

\bibitem{li2020small}
Ximin Li, Xiaodong Wei, and Xiaowei Qin,
\newblock ``Small-footprint keyword spotting with multi-scale temporal
  convolution,''
\newblock {\em Proc. Interspeech}, pp. 1987--1991, 2020.

\bibitem{choi2019temporal}
Seungwoo Choi, Seokjun Seo, Beomjun Shin, Hyeongmin Byun, Martin Kersner,
  Beomsu Kim, Dongyoung Kim, and Sungjoo Ha,
\newblock ``Temporal convolution for real-time keyword spotting on mobile
  devices,''
\newblock {\em Proc. Interspeech}, pp. 3372--3376, 2019.

\bibitem{rybakov2020streaming}
Oleg Rybakov, Natasha Kononenko, Niranjan Subrahmanya, Mirko Visontai, and
  Stella Laurenzo,
\newblock ``Streaming keyword spotting on mobile devices,''
\newblock {\em arXiv preprint arXiv:2005.06720}, 2020.

\bibitem{mo2020neural}
Tong Mo, Yakun Yu, Mohammad Salameh, Di~Niu, and Shangling Jui,
\newblock ``Neural architecture search for keyword spotting,''
\newblock {\em Proc. Interspeech}, pp. 1982--1986, 2020.

\bibitem{zhang2021autokws}
Bo~Zhang, Wenfeng Li, Qingyuan Li, Weiji Zhuang, Xiangxiang Chu, and Yujun
  Wang,
\newblock ``Autokws: Keyword spotting with differentiable architecture
  search,''
\newblock in {\em ICASSP 2021-2021 IEEE International Conference on Acoustics,
  Speech and Signal Processing (ICASSP)}. IEEE, 2021, pp. 2830--2834.

\bibitem{desplanques2020ecapa}
Brecht Desplanques, Jenthe Thienpondt, and Kris Demuynck,
\newblock ``Ecapa-tdnn: Emphasized channel attention, propagation and
  aggregation in tdnn based speaker verification,''
\newblock {\em arXiv preprint arXiv:2005.07143}, 2020.

\bibitem{chung2020defence}
Joon~Son Chung, Jaesung Huh, Seongkyu Mun, Minjae Lee, Hee~Soo Heo, Soyeon
  Choe, Chiheon Ham, Sunghwan Jung, Bong-Jin Lee, and Icksang Han,
\newblock ``In defence of metric learning for speaker recognition,''
\newblock {\em arXiv preprint arXiv:2003.11982}, 2020.

\end{thebibliography}

\end{document}